\documentclass[aps,pra,nofootinbib,preprint,amsmath,amssymb,floatfix]{revtex4}
\usepackage{color}
\usepackage{graphicx}

\begin{document}

\newcommand{\tc}{\textcolor}
\newcommand{\g}{blue}
\newcommand{\ve}{\varepsilon}
\title{Viscous accelerating universe with non-linear and logarithmic equation of state fluid}         

\author{I. Brevik$^1$, A. N. Makarenko$^2$ and A. V. Timoshkin$^{2,3,4}$ }      
\affiliation{$^1$Department of Energy and Process Engineering, Norwegian University of Science and Technology, N-7491 Trondheim, Norway}
\affiliation{$^2$Tomsk State Pedagogical University, Kievskaja Street, 60, 634061 Tomsk, Russia}
\affiliation{$^3$Tomsk State University of Control Systems and Radio Electronics, Lenin Avenue, 40, 634050 Tomsk, Russia}
\affiliation{$^4$National Research Tomsk State University,  Lenin Avenue, 36, 634050 Tomsk , Russia}

\begin{abstract}
We describe the accelerated expansion of the late-time universe using a generalized equation of state (EoS) when account is taken of bulk viscosity. We assume a homogeneous and isotropic Friedmann-Robertson-Walker spacetime. Solutions of the gravitational equations for dark energy are obtained in implicit form. Characteristic properties of the Universe evolution in the presence of the viscosity effects are discussed. Finally, the dynamics of the accelerated expansion of the viscous universe is discussed, on the basis of a modified logarithmic-corrected EoS.

\end{abstract}
\maketitle

\bigskip
\section{Introduction}
\label{secintro}
As is known, the explanation of the physical origin of the acceleration of the early and the later universe is a fundamental problem in contemporary physics. The discovery of the accelerated expansion has led to the emergence of new theoretical models of dark energy. The cosmic acceleration can be explained within the framework of scalar-tensor theories or via dark energy weekly interacting with ordinary matter \cite{1,2}, or via a modification of gravity theory \cite{3,4}. An extensive literature exists for the description of the universe using cosmological models for the dark energy fluid satisfying generalized EoS in modified theories of gravity. The obvious motivation for these attempts has been to be able to explain the acceleration in a natural way, and to predict the future evolution. Central elements in these theories are the presence of singularities in the cosmic fluid. There are various scenarious for the evolution of the accelerating universe. Notably, the final stage may be in the form of a Big Rip \cite{5,6,7}, a Little Rip \cite{8,9,10,11,12,13,14}, a Pseudo Rip \cite{15}, a Quasi Rip \cite{16}, or a bounce cosmology \cite{17,18}.

The universe with a viscous dark fluid was studied in Refs.~\cite{19,20,21,22,23,24,25,26,27,28}. It is notable that viscous fluids may be considered as a particular class of generalized fluids; cf. Refs.~\cite{29,30}. In a cosmological context the inclusion of viscosity greatly extends the versatility of the theory. When considering turbulence, for instance, one has to consider the effect of viscosity \cite{31}. In the violent motions of the cosmic fluid approaching a future singularity, it is reasonable that turbulent effects may occur \cite{32}. A viscous dark fluid can be considered as a model of modified gravity of $f(R)$ type \cite{33}. When the fluid is spatially isotropic, only the bulk viscosity comes into play. It turns out that this viscosity plays an important role in the future, near the Big Rip singularity \cite{34,35}, or in the so-called type II, type III, or type IV singularities \cite{36,37,38} according to the Nojiri-Odintsov-Tsujikawa classification. These types correspond to cases where one or more of the relevant physical quantities tend to infinity, in a finite time. Even in the very early universe, in the inflationary period, viscosity has been proposed in order to account for a calm exit from inflation. Thus a two-component viscous fluid has been successfully applied to describe an inflationary universe \cite{39}. Some examples inhomogeneous viscous coupled fluids were considered in \cite {40, 41}. In view of the many facets of viscous cosmology, it is quite natural that a large number of papers have appeared about this theme (cf., for example, the review paper \cite{42}).

In the present paper we will investigate the dynamical evolution of the late-time universe via a nonlinear inhomogeneous EoS for the dark energy fluid, taking the fluid to be viscous. We will examine in detail various forms of this inhomogeneous function, as well as the bulk viscosity, in the modified EoS and investigate how the universe evolution becomes affected by the equation of state. In particular, we will study the dynamical evolution of the late-time universe when applying an EoS with logarithmic terms in the energy density.

\section{Late-time university models with viscosity, from a modified equation of state}

We will study a viscous dark fluid in standard Einstein-Hilbert gravity, writing the Friedmann equation for the Hubble parameter as
\begin{equation}
\rho=\frac{3}{k^2}H^2. \label{1}
\end{equation}
Here $\rho$ is the energy density, $H=\dot{a}(t)/a(t)$ is the Hubble parameter with $a(t)$ the scale factor, $k^2=8\pi G$ with $G$ denoting Newton's gravitational constant, and a dot denotes a derivative with respect to cosmic time $t$.

We consider a spatially flat Friedmann-Robertson-Walker spacetime with metric
\begin{equation}
ds^2=-dt^2+a^2(t)\delta_{ij}\, dx^idx^j. \label{2}
\end{equation}
We will describe the behavior of the supposed homogeneous and isotropic fluid on large scales, at late times, when the EoS is nonlinear and inhomogenoeous,
\begin{equation}
p= \omega (\rho, t)\rho +f(\rho)+\zeta (H,t), \label{3}
\end{equation}
where $p$ is the pressure, $\omega$ the thermodynamic parameter depending on $\rho$ and $t$, $f(\rho)$ is an arbitrary function, and $\zeta (H,t)$ is a viscosity function depending on $H$ and $t$. As thermodynamics requires the viscosities (bulk and shear) to be positive, we shall require the viscosity function to be positive, $\zeta(H, t)>0$. It is known that, to a very good precision, an ideal fluid model with an inhomogeneus equation of state can satisfy the recent observed data \cite{31}.

Next, we will consider various cosmological models of a dark fluid, specifying the viscosity properties.

\subsection{Quadratic EoS with power-law viscosity}
We take the thermodynamic parameter, the nonlinear function, and the bulk viscosity, in the EoS equation (\ref{3}) to be of the form
\begin{equation*}
\omega (\rho, t)=b,
\end{equation*}
\begin{equation}
f(\rho)=c\rho^2+d\dot{\rho}^2, \label{4}
\end{equation}
\begin{equation*}
\zeta(H,t)=\xi_1(t)(3H)^n,
\end{equation*}
where $b,c,d$ are positive constants. We will choose $n=3$ in the viscosity function, and choose $\xi_1(t)=\tau$, where $\tau$ is a positive constant. Then, the viscosity function $\zeta(H,t)=27\tau H^3$ becomes a cubic function of the Hubble parameter. This form of the Hubble parameter is widely used in viscous cosmology \cite{43}. Using the Friedmann equation, Eq.~(\ref{3}) gets the following form,
\begin{equation}
p=b\rho +(c+9\tau k^4)\rho^2+d\dot{\rho}^2. \label{5}
\end{equation}
Let us consider the scenario in which $\rho$ fulfils the standard conservation equation
\begin{equation}
\dot{\rho}+3H(\rho+p)=0. \label{6}
\end{equation}
Hence, inserting (\ref{5}) into (\ref{6}) and using the Friedmann equation we obtain
\begin{equation}
d_1\sqrt{\rho}\,\dot{\rho}^2+\dot{\rho}+\bar{\tau}\rho^{5/2} + b_1\rho^{3/2}=0, \label{7}
\end{equation}
where $d_1=\sqrt{3}\,kd, \, \bar{\tau}=\sqrt{3}\, k(c+9\tau k^4)$, and $b_1=\sqrt{3}\, k(b+1)$.

Let us examine the case when $b=-1$. Then the solution of (\ref{7}) takes the form
\begin{equation}
\frac{2}{3}\sqrt{\frac{d_1}{\bar{\tau}}}
\int \frac{d\tilde{\rho}}{\pm \sqrt{1-\tilde{\rho}^2}-1}=t+C, \label{8}
\end{equation}
where $\tilde{\rho}=2\sqrt{d_1\bar{\tau}}\,\rho^{3/2}$ is a dimensionless energy density, and $C$ is an arbitrary constant.

Taking the positive sign in (\ref{8}) we find the following solution,
\begin{equation}
\tilde{\rho}^{-1}(1+\sqrt{1-\tilde{\rho}^2})
 + \arcsin \tilde{\rho} = \frac{3}{2}\sqrt{\frac{\bar{\tau}}{d_1}}\, t + \tilde{C}, \label{9}
\end{equation}
where $\tilde{C}=\frac{3}{2}\sqrt{\frac{\bar{\tau}}{d_1}}\, C$.

The derivative of $\tilde{\rho}$ is
\begin{equation}
\dot{\tilde{\rho}}(t)=-\frac{3}{2}\sqrt{\frac{\bar{\tau}}{d_1}}\, \frac{\tilde{\rho}^2}{1+\sqrt{1-\tilde{\rho}^2}}. \label{10}
\end{equation}
This is negative, corresponding to a decreasing energy density.

If we on the other hand choose the negative sign in Eq.~(\ref{8}) we get as solution of Eq.~(\ref{7})
\begin{equation}
\frac{\tilde{\rho}}{1+\sqrt{1-\tilde{\rho}^2}}-\arcsin \tilde{\rho}=\frac{3}{2}\sqrt{\frac{\bar{\tau}}{d_1}}\, t +\tilde{C}, \label{11}
\end{equation}
and the derivative of $\tilde{\rho}$ becomes
\begin{equation}
\dot{\tilde{\rho}}=\frac{3}{2}\sqrt{\frac{\bar \tau}{d_1}}
(1+\sqrt{1-\tilde{\rho}^2}). \label{12}
\end{equation}
This is positive, the energy density increases as the universe expands.

\subsection{Fractional EoS with constant viscosity}

Let us choose the following functions in Eq.~(\ref{3}),
\begin{equation*}
\omega (\rho, t)=b,
\end{equation*}
\begin{equation}
f(\rho)=c\sqrt{\rho}+\frac{d}{\dot{\rho}}, \label{13}
\end{equation}
\begin{equation*}
\zeta(H,t)=\zeta_0,
\end{equation*}
where $\zeta_0$ is the constant viscosity function. In this case the EoS has the following form,
\begin{equation}
b=b\rho +c\sqrt{\rho} +\frac{d}{\dot{\rho}}+3\zeta_0H. \label{14}
\end{equation}
If we choose the value $b=-1$, the equation of motion (\ref{3}) becomes
\begin{equation}
\dot{\rho}^2+\bar{\zeta}_0\rho \dot{\rho}+d_1\sqrt{\rho}=0, \label{15}
\end{equation}
where $\bar{\zeta}_0=\sqrt{3}\, k(c+\sqrt{3}\, k \zeta_0)$ and $d_1=\sqrt{3}\, kd$.

The solution of (\ref{15}) is
\begin{equation}
\rho^{3/4}\sqrt{ \left(\frac{\bar{\zeta}_0}{2}\right)^2 \rho^{3/2}-d_1}
 -\frac{2d_1}{\bar{\zeta}_0}\ln \left( \sqrt{\rho^{3/2}
-\frac{4d_1}{\bar{\zeta}_0}}
 +\rho^{3/4}\right)
  -\frac{\bar{\zeta}_0}{2}\rho^{3/2}=C-\frac{3}{2}d_1t, \label{16}
\end{equation}
and the derivative of $\rho$ is
\begin{equation}
\dot{\rho}=\rho^{1/4}\left( \sqrt{\left( \frac{\bar{\zeta}_0}{2}\right)^2 \rho^{3/2}-\sqrt{3}\, kd} -\frac{\bar{\zeta}_0}{2}\rho^{3/4}\right). \label{17}
\end{equation}
If $d<0$ then $\dot{\rho}>0$; the energy density increases and the universe expands. Vice versa, if $d>0$ then $\dot{\rho}<0$ and the energy density decreases.

\section{Logarithmic EoS with viscosity}

Let us now investigate the dark energy based upon a logarithmic-corrected power-law fluid. We assume the EoS to have the form \cite{44}
\begin{equation}
p=A\left( \frac{\rho}{\rho_*}\right)^{-l}\, \ln \left( \frac{\rho}{\rho_*}\right), \label{18}
\end{equation}
where $\rho_*$ is a reference density. Following \cite{45} it will be identified with the Planck density, $\rho_p=c^5/\hbar G^2 \approx 5.16\times 10^{99}~$g/m$^3$. In this notation $A$ represents the logotropic temperature, $l= -\frac{1}{6}-\gamma_G$, where $\gamma_G$ is the dimensional Gruneisen parameter. This parameter
  is a free parameter in the theory of the homogeneous and isotropic universe. For the case $l=0$ we obtain the EoS for the logotropic cosmological model \cite{45}. The logarithmic-corrected power-law fluid possesses properties analogous to the properties of isotropic crystalline solids, where the pressure can be negative. This formalism allows us to model, and explain, the accelerating expansion of the late universe in terms of the logotropic dark fluid. In order to obtain a more detailed description of the evolution of the late universe, we will include, together with the EoS (\ref{18}), a bulk viscosity term.

  \subsection{Viscosity with a constant function $\xi_1(t)$}

  We will assume that the universe is homogeneous, isotropic, and spatially flat. We modify the set of equations (\ref{4}), taking into account the logarithmic for $p$ in (\ref{18}), to obtain
  \begin{equation*}
  \omega(\rho,t)=b,
  \end{equation*}
  \begin{equation}
  f(\rho)=A\left( \frac{\rho}{\rho_*}\right)^{-l}\ln \left( \frac{\rho}{\rho_*}\right) +B\rho^\alpha, \label{19}
  \end{equation}
  \begin{equation*}
  \zeta(H,t)=3\gamma H.
  \end{equation*}
  here $\alpha, \gamma$ are positive constants. Then the bulk viscosity function is a linear function of the Hubble parameter. The case of $B=0$ was considered in Ref.~\cite{46}.

  The EoS (\ref{3}) gets the form
  \begin{equation}
  p=b\rho +A\left( \frac{\rho}{\rho_*}\right)^{-l}\ln \left(\frac{\rho}{\rho_*}\right) +B\rho^\alpha -3k^2 \gamma \rho. \label{20}
  \end{equation}
  We will now study the case when $b=-1$ and $\alpha =1/2$. Then we obtain a modified gravitational equation for the viscous dark fluid,
  \begin{equation}
  \dot{\rho}+3H\left[ A\left(\frac{\rho}{\rho_*}\right)^{-l}\ln \left(\frac{\rho}{\rho_*}\right) +B\sqrt{\rho}-3\gamma k^2\rho \right]=0. \label{21}
  \end{equation}
  In the high energy regime $(\rho >\rho_*)$, and for $l=-1$, we obtain from (\ref{21}) using the Friedmann equation
  \begin{equation}
  \dot{H}+\frac{3}{2\rho_*}(A-3\gamma k^2\rho_*)H^2+\frac{\sqrt{3}}{2}BkH-\frac{1}{2}Ak^2=0. \label{22}
  \end{equation}
  The solution of this equation is
  \begin{equation}
  H(t)=d\,\frac{e^{\tilde{\theta}t/2}+Ce^{-\tilde{\theta}t/2}}{ e^{\tilde{\theta}t/2}-Ce^{-\tilde{\theta}t/2}} -c, \label{23}
  \end{equation}
  where
  \[ c=\frac{Bk\rho_*}{2\sqrt{3}(A-3\gamma k^2\rho_*)}, \quad d=\frac{k\rho_*\sqrt{B^2+4(A-3\gamma k^2\rho_*)}}{2\sqrt{3}(A-3\gamma k^2\rho_*)}, \quad \tilde{\theta}=\frac{\rho_*}{3d(A-3\gamma k^2\rho_*)},       \]
  and $C$ is an arbitrary constant. It is thus seen that the Hubble parameter $H(t)$ diverges for $t\rightarrow 0$ and a Big Bang singularity occurs. If we in addition take $C=1$ for simplicity, we get for the scale factor
  \begin{equation}
  a(t)=a_0e^{-ct}\left( \sinh \frac{1}{2}\tilde{\theta}t\right)^{2d/\tilde{\theta}}, \label{24}
  \end{equation}
  where $a_0$ is an integration constant. Its second derivative is
  \begin{equation}
  \ddot{a}(t)=a(t)\left[ \left(d\coth \frac{1}{2}\tilde{\theta}t -c\right)^2
  -\frac{d\tilde{\theta}}{2\sinh^2 \frac{1}{2}\tilde{\theta}t} \right]. \label{25}
  \end{equation}
  We obtain $\ddot{a}(t)=0$ at
  \begin{equation}
  t_0=\sqrt{3}k\sqrt{B^2+4(A-3\gamma k^2\rho_*)}\left[ \frac{1}{2}\ln \frac{2(A-3\gamma k^2\rho_*)}{\rho_*k^2}+\ln \frac{\sqrt{1+2\rho_*k}-1}{\sqrt{B^2+4(A-3\gamma k^2\rho_*)}-B}\right], \label{26}
  \end{equation}
  where $A-3\gamma k^2\rho_*>0$.

  If $B\neq 0$ the second derivative of the scale factor is positive, so that the universe expands with acceleration (as is the case of the universe at present). If $B=0$ for earlier time $0<t<t_0$, the second derivative is negative. In this case the universe is expanding, though  decelerating, while for $t>t_0$ we get an expanding universe with accelerated expansion.

  From (\ref{23}) we find that
  \begin{equation}
  \dot{H}(t)=-\frac{b\tilde{\theta}}{2\sinh^2 \frac{1}{2}\tilde{\theta}t}, \label{27}
  \end{equation}
showing that since $\dot{H}(t)<0$ the universe does not super-accelerate. Thus, this model can describe the evolution of an universe with transition from a decelerating to an accelerating era. Note that we have not been considering a cosmological constant explicitly; thus the calculated behavior is a consequence of the dynamics of our model only.

\subsection{Viscosity with a linear time-dependent function $\xi_1(t)$}

Let us assume the function in (\ref{4}) to be linear in time,
\begin{equation}
\xi_1(t)=mt+r, \label{28}
\end{equation}
$m$ and $r$ being arbitrary parameters. Then in the case $n=1$ the viscosity function has the following simple form,
\begin{equation}
\zeta(H,t)=3H(mt+r). \label{29}
\end{equation}
The the equation of state has the form
\begin{equation}
p=b\rho+A\left(\frac{\rho}{\rho_*}\right)^{-l}\ln \left( \frac{\rho}{\rho_*}\right) +B\rho^\alpha -3k^2(mt+r)\rho. \label{30}
\end{equation}
We will study the case where $b=-1$ and $\alpha=1$. Then, inserting (\ref{30}) into (\ref{6}) we obtain the modified gravitational equation of motion,
\begin{equation}
\dot{\rho}+3H\left[ A\left(\frac{\rho}{\rho_*}\right)^{-l}\ln \left(\frac{\rho}{\rho_*}\right) +B\rho -3k^2(mt+r)\rho \right]=0. \label{31}
\end{equation}
In the case $l=-1$, using the Friedmann equation, we can rewrite (\ref{31}) as
\begin{equation}
\dot{H}+(a+b_1t)H^2 -h=0, \label{32}
\end{equation}
where
\begin{equation*}
a=\frac{3}{2}\left( B+\frac{A}{\rho_*}-3k^2r\right), \quad b_1=-\frac{27}{4}k^2m, \quad h=\frac{1}{2}Ak^2.
\end{equation*}
The solution of (\ref{32}) is
\begin{equation}
H(t)=\frac{u'(t)}{f(t)u(t)}, \label{33}
\end{equation}
with $f(t)=a+b_1t$.

Let us consider the limit $t\rightarrow \infty$. Then
\begin{equation}
u(t)=C\eta(t)Z_{1/3}\left( \frac{2}{3hb_1}\eta^{3/2}(t)\right), \label{34}
\end{equation}
where $Z_\nu(t)$ is a modified Bessel function, $\eta(t)=hf(t)$, and $C$ is an arbitrary constant. When $t \rightarrow \infty$, then $\eta \rightarrow \infty$, so that for fixed $\nu$ and increasing values of $t$ the function $Z_\nu$ increases. For $x\rightarrow \infty$ we have $Z_\nu(x) \approx \frac{e^x}{\sqrt{2\pi x}}$ so that in this approximation
\begin{equation}
u(t)=\tilde{C}\eta^{1/4}\exp\left( \frac{2}{3hb_1}\eta^{3/2}\right), \label{35}
\end{equation}
where $\tilde{C}=\frac{3}{4}\sqrt{\frac{b_1}{\pi}}C$. In view of this we can write the solution (\ref{33}) for $H(t)$ as
\begin{equation}
H(t)=\tilde{C}\frac{1}{\sqrt{\eta}}\left( \frac{hb_1}{4\eta^{3/2}}+1\right). \label{36}
\end{equation}
Notice that the Hubble parameter $H (t)$ diverges at a finite time $t_0=\frac{1}{3k^2m}\left( B+\frac{A}{\rho_*}-3r\right)$, meaning that a Big Rip type singularity appears.

The time derivative of $H(t)$ becomes
\begin{equation}
\dot{H}(t)=-\frac{1}{2}\tilde{C}\,\frac{hb_1}{\eta^{3/2}}\left( \frac{hb_1}{\eta^{3/2}}+1\right). \label{37}
\end{equation}
If the parameters $h$ and $b$ are positive, then the derivative of the Hubble parameter is negative. Thus, because $\dot{H}(t)<0$, the universe does not super-accelerate.

Using the solution (\ref{36}), we calculate the scale factor
\begin{equation}
a(t)=a_0\exp\left( \frac{3\tilde{C}}{2\sqrt{\pi b_1}}
\sqrt{a+b_1t}\right), \label{38}
\end{equation}
with $a_0$ an integration constant. Its second derivative is
\begin{equation}
\ddot{a}(t)=a(t)\frac{\tilde{C}}{h(a+b_1t)}\left( 1-\frac{1}{2}\frac{\sqrt{h}\,b_1}{\sqrt{a+b_1t}}\right). \label{39}
\end{equation}
Hence, $\ddot{a}(t)=0$ at $t_0'=\frac{hb_1}{4}-\frac{a}{b_1}$. Thus, if $A>0$ we obtain $\ddot{a}(t)<0$ for $t<t_0'$, leading to a decelerated expansion, while for $t>t_0'$ we have $\ddot{a}(t)>0$ and the universe transits to a late-time accelerated era. In the limiting case $t\rightarrow \infty$ this model can describe a universe evolution containing a transition from a decelerating to an accelerating era.

\section{Conclusions}

We have studied a dark energy model of the late-time universe, bases on a nonlinear inhomogeneous EoS in the presence of a bulk viscosity. We assumed a homogeneous and isotropic spatially flat Friedmann-Robertson-Walker spacetime.Various classes were studied, corresponding to different nonlinear functions and viscosity forms. This formalism allows one to model and explain the accelerating expansion of the late universe in terms of parameters contained in the modified dark fluid EoS. We obtained an accelerating expansion of the late universe, with transition from a decelerating phase to an accelerating one, without involving a cosmological constant explicitly. We investigated also the dynamical evolution of the late-time universe via dark energy fluids having a logarithmic equation of state. We examined various forms of the dark energy equation of state in detail,  and analyzed how the universe evolution depends on the choice of parameters in the corresponding EoS. We showed that the dynamics of the model was the  main factor for the behaviour of the universe.

Finally let us mention the pioneering papers of Elizalde {\it et al.} \cite{47}, and of Cognola {\it et al.} \cite{48}, which are closely related to the main theme of our work, although from a different perspective.


\end{document}